\documentclass[10pt, conference, letterpaper]{IEEEtran}
\pdfoutput=1

\usepackage{graphicx}
\usepackage[font=footnotesize]{caption}
\usepackage[labelformat=simple]{subcaption}
\usepackage{epstopdf}
\graphicspath{{figs/}}
\usepackage{amsthm}
\usepackage{amssymb}
\usepackage{amsmath}
\usepackage{verbatim}
\usepackage{cite}
\usepackage{url}
\usepackage{setspace}

\newcommand{\eg}{\emph{e.g.}}
\newcommand{\ie}{\emph{i.e.}}
\newcommand{\etal}{\emph{et al.}}

\newcommand{\diff}{\mathrm{d}}
\newcommand{\LL}{\mathcal{L}}

\newcommand{\h}{\mathbf{h}}

\newcommand{\E}{\mathbb{E}}

\newcommand{\Prob}{\mathbb{P}}
\newcommand{\reals}{\mathbb{R}}

\newtheorem{theorem}{Theorem}

\title{A Utility Optimization Approach\\ to Network Cache Design}
{\author{Mostafa Dehghan$^1$, Laurent Massoulie$^2$, Don Towsley$^1$, Daniel Menasche$^3$, and Y. C. Tay$^4$\\

$^1$University of Massachusetts Amherst,
$^2$Microsoft Research - INRIA Joint Center \\
$^3$Federal University of Rio de Janeiro, 
$^4$National University of Singapore \\
{\tt \{mdehghan, towsley\}@cs.umass.edu}, {\tt laurent.massoulie@inria.fr}\\ {\tt sadoc@dcc.ufrj.br}, {\tt dcstayyc@nus.edu.sg}
}

\begin{document}

\maketitle

\thispagestyle{plain}
\pagestyle{plain}

\begin{abstract}
In any caching system, the admission and eviction policies determine which contents are added and removed from a cache when a miss occurs. Usually, these policies are devised so as to mitigate staleness and increase the hit probability. Nonetheless, the utility of having a high hit probability can vary across contents. This occurs, for instance, when service level agreements must be met, or if certain contents are more difficult to obtain than others.
In this paper, we propose utility-driven caching, where we associate with each content a utility, which is a function of the corresponding content hit probability.
We formulate optimization problems where the objectives are to maximize the sum of utilities over all contents. These problems differ according to the stringency of the cache capacity constraint.
Our framework enables us to reverse engineer classical replacement policies such as LRU and FIFO, by computing the utility functions that they maximize.
We also develop online algorithms that can be used by service providers to implement various caching policies based on arbitrary utility functions.
\end{abstract}
\section{Introduction}
The increase in data traffic over past years is predicted to continue more aggressively, with global Internet traffic in 2019 estimated to reach 64 times of its volume in 2005~\cite{cisco14}.
The growth in data traffic is recognized to be due primarily to streaming of video on-demand content over cellular networks.
However, traditional methods such as increasing the amount of spectrum or deploying more base stations are not sufficient to cope with this predicted traffic increase~\cite{Andrews12, Golrezaei12}. Caching is recognized, in current and future Internet architecture proposals, as one of the most effective means to improve the performance of web applications.
By bringing the content closer to users, caches greatly reduce network bandwidth usage, server load, and perceived service delays~\cite{borst10}.


Because of the trend for ubiquitous computing, creation of new content publishers and consumers, the Internet is becoming an increasingly heterogeneous environment where different content types have different quality of service requirements, depending on the content publisher/consumer.
Such an increasing diversity in service expectations advocates the need for content delivery infrastructures with service differentiation among different applications and content classes.
Service differentiation not only induces important technical gains, but also provides significant economic benefits~\cite{feldman02}.
Despite a plethora of research on the design and implementation of \emph{fair} and \emph{efficient} algorithms for differentiated bandwidth sharing in communication networks, little work has focused on the provision of multiple levels of service in network and web caches.
The little available research has focused on designing controllers for partitioning cache space~\cite{ko03, lu04}, biased replacement policies towards particular content classes~\cite{kelly99}, or using multiple levels of caches~\cite{feldman02}. These techniques either require additional controllers for fairness, or inefficiently use the cache storage.

Moreover, traditional cache management policies such as LRU treat different contents in a strongly coupled manner that makes it difficult for (cache) service providers to implement differentiated services, and for content publishers to account for the valuation of their content delivered through content distribution networks.
In this paper, we propose a utility-driven caching framework, where each content has an associated utility and content is stored and managed in a cache so as to maximize the aggregate utility for all content. Utilities can be chosen to trade off user satisfaction and cost of storing the content in the cache.
We draw on analytical results for time-to-live (TTL) caches~\cite{Nicaise14b}, to design caches with ties to utilities for individual (or classes of) contents.
Utility functions also have implicit notions of fairness that dictate the time each content stays in cache.
Our framework allows us to develop \emph{online} algorithms for cache management, for which we prove achieve optimal performance. Our framework has implications for distributed pricing and control mechanisms and hence is well-suited for designing cache market economic models.

Our main contributions in this paper can be summarized as follows:
\begin{itemize}
\item We formulate a utility-based optimization framework for maximizing aggregate content publisher utility subject to buffer capacity constraints at the service provider.
We show that existing caching policies, \eg\ LRU, LFU and FIFO, can be modeled as utility-driven caches within this framework.
\item By reverse engineering the LRU and FIFO caching policies as utility maximization problems, we show how the \emph{characteristic time}~\cite{Che01} defined for these caches relates to the Lagrange multiplier corresponding to the cache capacity constraint.
\item We develop online algorithms for managing cache content, and prove the convergence of these algorithms to the optimal solution using Lyapunov functions.
\item We show that our framework can be used in revenue based models where content publishers react to prices set by (cache) service providers without revealing their utility functions.
\item We perform simulations to show the efficiency of our online algorithms using different utility functions with different notions of fairness.
\end{itemize}

The remainder of the paper is organized as follows. We review related work in the next section. Section~\ref{sec:model} explains the network model considered in this paper, and
Section~\ref{sec:opt} describes our approach in designing utility maximizing caches. In Section~\ref{sec:fairness} we elaborate on fairness implications of utility functions, and in Section~\ref{sec:reverse}, we derive the utility functions maximized by LRU and FIFO caches. In Section~\ref{sec:online}, we develop online algorithms for implementing utility maximizing caches. We present simulation results in Section~\ref{sec:simulation}, and discuss prospects and implications of the cache utility maximization framework in Section~\ref{sec:discussion}. Finally, we conclude the paper in Section~\ref{sec:conclusion}.

\section{Related Work}
\subsection{Network Utility Maximization}

Utility functions have been widely used in the modeling and control of computer networks, from stability analysis of queues to the study of fairness in network resource allocation; see~\cite{srikant13, neely10} and references therein. Kelly~\cite{kelly97} was the first to formulate the problem of rate allocation as one of achieving maximum
aggregate utility for users, and describe how network-wide optimal rate allocation can be achieved by having individual users control their transmission rates.
The work of Kelly~\etal~\cite{kelly98} presents the first mathematical model and analysis of the behavior of congestion control algorithms for general topology networks.
Since then, there has been extensive research in generalizing and applying Kelly's \emph{Network Utility Maximization} framework to model and analyze various network protocols and architectures. This framework has been used to study problems such as network routing~\cite{tassiulas92}, throughput maximization~\cite{eryilmaz07}, dynamic power allocation~\cite{neely03} and scheduling in energy harvesting networks~\cite{huang13}, among many others. Ma and Towsley~\cite{Ma15} have recently proposed using utility functions for the purpose of designing contracts that allow service providers to monetize caching.


\subsection{Time-To-Live Caches}
TTL caches, in which content eviction occurs upon the expiration of a timer, have been employed
since the early days of the Internet with the Domain Name System (DNS) being an important application~\cite{Jung03}. More recently, TTL caches have regained popularity, mostly due to admitting a general approach in the analysis of caches that can also be used to model replacement-based caching policies such as LRU. The connection between
TTL caches and replacement-based (capacity-driven) policies was first established for the LRU policy by Che~\etal~\cite{Che01} through the notion of cache \emph{characteristic time}. The characteristic time was theoretically justified and extended to other caching policies such as FIFO and RANDOM~\cite{Fricker12}. 
This connection was further confirmed to hold for more general arrival models than Poisson processes~\cite{Bianchi13}. Over the past few years, several exact and approximate analyses have been proposed for modeling single caches in isolation as well as cache networks using the TTL framework~\mbox{\cite{Nicaise14, Berger14}}.

In this paper, we use TTL timers as \emph{tuning knobs} for individual (or classes of) files to control the utilities observed by the corresponding contents,
and to implement \emph{fair} usage of cache space among different (classes of) contents.
We develop our framework based on two types of TTL caches described in the next section.

%
%
\section{Model}
\label{sec:model}
Consider a set of $N$ files, and a cache of size $B$. We use the terms file and content interchangeably in this paper.
Let $h_i$ denote the hit probability for content $i$.
Associated with each content, $i=1,\ldots, N$, is a utility function $U_i:[0,1] \rightarrow \reals$ that represents the ``satisfaction'' perceived by observing hit probability $h_i$.
$U_i(\cdot)$ is assumed to be increasing, continuously differentiable, and strictly concave.
Note that a function with these properties is invertible. We will treat utility functions that do not satisfy these constraints as special cases.

\subsection{TTL Caches}
In a TTL cache, each content is associated with a timer~$t_i$.  Whenever a cache miss to content $i$ occurs, content $i$ is stored in the cache and its timer is set to $t_i$.
Timers decrease at constant rate, and a content is evicted from cache when its timer reaches zero.
We can adjust the hit probability of a file by controlling the time a file is kept in cache.

There are two TTL cache designs:
\begin{itemize}
\item Non-reset TTL Cache: TTL is only set at cache misses, \ie~TTL is not reset upon cache hits.
\item Reset TTL Cache: TTL is set each time the content is requested.
\end{itemize}
Previous work on the analysis of TTL caches~\cite{Nicaise14} has shown that the hit probability of file $i$ for these two classes of non-reset and reset TTL caches can be expressed as 
\begin{equation}
\label{eq:hit_non_reset}
h_i = 1 - \frac{1}{1 + \lambda_i t_i},
\end{equation}
and
\begin{equation}
\label{eq:hit_reset}
h_i = 1 - e^{-\lambda_i t_i},
\end{equation}
respectively, where requests for file $i$ arrive at the cache according to a Poisson process with rate $\lambda_i$.
Note that depending on the utility functions, different (classes of) files might have different or equal TTL values.

\section{Cache Utility Maximization}
\label{sec:opt}
In this section, we formulate cache management as a utility maximization problem.  We introduce two formulations, one where the buffer size introduces a hard constraint and a second where it introduces a soft constraint.

\subsection{Hard Constraint Formulation}
We are interested in designing a cache management policy that optimizes the sum of utilities over all files, more precisely,
\begin{align}
\label{eq:opt}
\text{maximize} \quad &\sum_{i=1}^{N}{U_i(h_i)} \notag\\
\text{such that} \quad &\sum_{i=1}^{N}{h_i} = B \\
& 0 \le h_i \le 1, \quad i=1, 2, \ldots, N. \notag
\end{align}
Note that the feasible solution set is convex and since the objective function is strictly concave and continuous, a unique maximizer, called the optimal solution, exists.  Also note that the buffer constraint is based on the {\em expected} number of files not exceeding the buffer size and not the total number of files.
Towards the end of this section, we show that the buffer space can be managed in a way such that the probability of \emph{violating} the buffer size constraint vanishes as the number of files and cache size grow large.

This formulation does not enforce any special technique for managing the cache content, and any strategy that can easily adjust the hit probabilities can be employed. We use the TTL cache as our building block because it provides the means through setting timers to control the hit probabilities of different files in order to maximize the sum of utilities.

Using timer based caching techniques for controlling the hit probabilities with $0 < t_i < \infty$ ensures that $0 < h_i < 1$, and hence, disregarding the possibility of $h_i = 0$ or $h_i = 1$, we can write the Lagrangian function as
\begin{align*}
\LL(\mathbf{h}, \alpha) &= \sum_{i=1}^{N}{U_i(h_i)}-\alpha\left[ \sum_{i=1}^{N}{h_i} - B\right] \\
&= \sum_{i=1}^{N}{\Big[ U_i(h_i)-\alpha h_i \Big]} + \alpha B,
\end{align*}
where $\alpha$ is the Lagrange multiplier. 

In order to achieve the maximum in $\LL(\mathbf{h}, \alpha)$, the hit probabilities should satisfy
\begin{equation}
\label{eq:drv}
\frac{\partial\LL}{\partial h_i} = \frac{\diff U_i}{\diff h_i} - \alpha = 0.
\end{equation}

Let $U'_i(\cdot)$ denote the derivative of the the utility function $U_i(\cdot)$, and define ${U'_i}^{-1}(\cdot)$ as its inverse function.
From~\eqref{eq:drv} we get
\[U'_i(h_i) = \alpha,\]
or equivalently
\begin{equation}
\label{eq:hu}
h_i = {U'_i}^{-1}(\alpha).
\end{equation}
Applying the cache storage constraint we obtain
\begin{equation}
\label{eq:c}
\sum_{i}{h_i} = \sum_{i}{{U'_i}^{-1}(\alpha)} = B,
\end{equation}
and $\alpha$ can be computed by solving the fixed-point equation given above.

As mentioned before, we can implement utility maximizing caches using TTL based policies.
Using the expression for the hit probabilities of non-reset and reset TTL caches given in~\eqref{eq:hit_non_reset} and~\eqref{eq:hit_reset},
we can compute the timer parameters $t_i$, once $\alpha$ is determined from~\eqref{eq:c}.
For non-reset TTL caches we obtain
\begin{equation}
\label{eq:non_reset_t}
t_i = -\frac{1}{\lambda_i}\Big(1 - \frac{1}{1 - {U'_i}^{-1}(\alpha)}\Big),
\end{equation}
and for reset TTL caches we get
\begin{equation}
\label{eq:reset_t}
t_i = -\frac{1}{\lambda_i}\log{\Big(1 - {U'_i}^{-1}(\alpha)\Big)}.
\end{equation}

\subsection{Soft Constraint Formulation}
\label{sec:soft}
The formulation in~\eqref{eq:opt} assumes a hard constraint on cache capacity.
In some circumstances it may be appropriate for the (cache) service provider to increase the available cache storage at some cost to the file provider
for the additional resources\footnote{One straightforward way of thinking this is to turn the cache memory disks on and off based on the demand.}.
In this case the cache capacity constraint can be replaced with a penalty function $C(\cdot)$ denoting the cost for the extra cache storage.
Here, $C(\cdot)$ is assumed to be a convex and increasing function.
We can now write the utility and cost driven caching formulation as
\begin{align}
\label{eq:opt_soft}
\text{maximize} \quad &\sum_{i=1}^{N}{U_i(h_i)} - C(\sum_{i=1}^{N}{h_i} - B) \\
\text{such that} \quad &0 \le h_i \le 1, \quad i=1,2,\ldots, N. \notag
\end{align}

Note the optimality condition for the above optimization problem states that
\[U'_i(h_i) = C'(\sum_{i=1}^{N}{h_i} - B).\]

Therefore, for the hit probabilities we obtain
\[h_i = {U'_i}^{-1}\Big(C'(\sum_{i=1}^{N}{h_i} - B)\Big),\]
and the optimal value for the cache storage can be computed using the fixed-point equation
\begin{equation}
\label{eq:elastic_B}
B^* = \sum_{i=1}^{N}{{U'_i}^{-1}\Big(C'(B^* - B)\Big)}.
\end{equation}

\subsection{Buffer Constraint Violations}
\label{sec:violation}
Before we leave this section, we address an issue that arises in both formulations, namely how to deal with the fact that there may be more contents with unexpired timers than can be stored in the buffer.  This occurs in the formulation of (\ref{eq:opt}) because the constraint is on the {\em average} buffer occupancy and in (\ref{eq:opt_soft}) because there is no constraint. Let us focus on the formulation in (\ref{eq:opt}) first.  Our approach is to provide a buffer of size $B(1+\epsilon )$ with $\epsilon > 0$, where a portion $B$ is used to solve the optimization problem and the additional portion $\epsilon B$ to handle buffer violations.  We will see that as the number of contents, $N$,  increases, we can get by growing $B$ in a sublinear manner, and allow $\epsilon$ to shrink to zero, while ensuring that content will not be evicted from the cache before their timers expire with high probability.  Let  $X_i$ denote whether content $i$ is in the cache or not; $P(X_i =1) = h_i $.  Now Let $\E\bigl[\sum_{i=1}^N X_i\bigr] = \sum_{i=1}^N h_i = B$. We write $B(N)$ as a function of $N$, and assume that $B(N) = \omega (1)$. 
\begin{theorem}
\label{thrm:violation}
For any $\epsilon > 0$
\[
 \Prob\bigl(\sum_{i=1}^N X_i \ge B(N)(1+\epsilon)\bigr) \le e^{-\epsilon^2 B(N)/3} .
\]
\end{theorem}
The proof follows from the application of a Chernoff bound.

Theorem~\ref{thrm:violation} states that we can size the buffer as $B(1+\epsilon)$ while using a portion $B$ as the constraint in the optimization.  The remaining portion, $\epsilon B$, is used to protect against buffer constraint violations. 
It suffices for our purpose that ${\epsilon^2 B(N) = \omega (1)}$.  This allows us to select $B(N) = o(N)$ while at the same time selecting $\epsilon = o(1)$. As an example, consider Zipf's law with $\lambda_i = \lambda/i^s$, $\lambda > 0$, $0 < s <1$,  $i=1,\ldots, N$ under the assumption that $\max{\{t_i\}} = t$ for some $t <\infty$.  In this case, we can grow the buffer as $B(N) = O(N^{1-s})$ while 
$\epsilon$  can shrink as $\epsilon = 1/N^{(1-s)/3}$.  Analogous expressions can be derived for $s \ge 1$.

Similar choices can be made for the soft constraint formulation.

\section{Utility Functions and Fairness}
\label{sec:fairness}
Using different utility functions in the optimization formulation~\eqref{eq:opt} yields different timer values for the files.
In this sense, each utility function defines a notion of fairness in allocating storage resources to different files.
In this section, we study a number of utility functions that have important fairness properties associated with them.
\subsection{Identical Utilities}
\label{sec:opt_identical}
Assume that all files have the same utility function, \ie\ $U_i(h_i) = U(h_i)$ for all $i$. Then, from~\eqref{eq:c} we obtain
\[\sum_{i=1}^{N}{{U'}^{-1}(\alpha)} = N {U'}^{-1}(\alpha) = B,\]
and hence
\[{U'}^{-1}(\alpha) = B/N.\]
Using~\eqref{eq:hu} for the hit probabilities we get
\[h_i = B/N, \quad \forall{i}.\]

Using a non-reset TTL policy, the timers should be set according to
\[t_i = \frac{B}{\lambda_i (N - B)},\]
while with a reset TTL policy, they must equal
\[t_i = -\frac{1}{\lambda_i}\log{\left(1-\frac{B}{N}\right)}.\]

The above calculations show that identical utility functions yield identical hit probabilities for all files. Note that the hit probabilities computed above do not depend on the utility function.

\subsection{$\boldsymbol{\beta}$-Fair Utility Functions}
Here, we consider the family of $\beta$-fair (also known as \emph{isoelastic}) utility functions given by
\[U_i(h_i) = \left\{ \begin{array}{ll}
 w_i\frac{h_i^{1-\beta}}{1-\beta} & \beta \ge 0, \beta \neq 1; \\
 & \\
 w_i \log{h_i} & \beta = 1,
 \end{array} \right. \]
where the coefficient $w_i \ge 0$ denotes the weight for file $i$.
This family of utility functions unifies different notions of fairness in resource allocation~\cite{srikant13}.
In the remainder of this section, we investigate some of the choices for $\beta$ that lead to interesting special cases.

\subsubsection{$\boldsymbol{\beta = 0}$}\hspace*{\fill} \\
With $\beta = 0$, we get $U_i(h_i) = w_i h_i$, and maximizing the sum of the utilities
corresponds to
\[\max_{h_i}{\sum_{i}{w_i h_i}}.\]

The above utility function defined does not satisfy the requirements for a utility function mentioned in Section~\ref{sec:model}, as it is not strictly concave.
However, it is easy to see that the sum of the utilities is maximized when
\[h_i = 1, i=1,\ldots, B  \quad \text{ and } \quad h_i = 0, i=B+1,\ldots, N,\]
where we assume that weights are sorted as ${w_1 \ge \ldots \ge w_N}$.
These hit probabilities indicate that the optimal timer parameters are
\[t_i = \infty, i=1,\ldots, B  \quad \text{ and } \quad t_i = 0, i=B+1,\ldots, N.\]

Note that the policy obtained by implementing this utility function with $w_i = \lambda_i$ corresponds to the Least-Frequently Used (LFU) caching policy,
and maximizes the overall throughput.

\subsubsection{$\boldsymbol{\beta = 1}$}\hspace*{\fill} \\
Letting $\beta = 1$, we get $U_i(h_i) = w_i \log{h_i}$,
and hence maximizing the sum of the utilities corresponds to
\[\max_{h_i}{\sum_{i}{w_i \log{h_i}}}.\]

It is easy to see that ${U'_i}^{-1}(\alpha) = w_i / \alpha$, and hence using~\eqref{eq:c} we obtain
\[\sum_{i}{{U'_i}^{-1}(\alpha)} = \sum_{i}{w_i} / \alpha = B,\]
which yields
\[\alpha = \sum_{i}{w_i} / B.\]
The hit probability of file $i$ then equals
\[h_i = {U'_i}^{-1}(\alpha) = \frac{w_i}{\sum_{j}{w_j}}B.\]

This utility function implements a \emph{proportionally fair} policy~\cite{kelly98}.
With $w_i = \lambda_i$, the hit probability of file $i$ is proportional to the request arrival rate $\lambda_i$.

\subsubsection{$\boldsymbol{\beta = 2}$}\hspace*{\fill} \\
With $\beta = 2$, we get $U_i(h_i) = -w_i/h_i$, and maximizing the total utility corresponds to
\[\max_{h_i}{\sum_{i}{\frac{-w_i}{h_i}}}.\]

In this case, we get ${U'_i}^{-1}(\alpha) = \sqrt{w_i} / \sqrt{\alpha}$, therefore
\[\sum_{i}{{U'_i}^{-1}(\alpha)} = \sum_{i}{\sqrt{w_i}} / \sqrt{\alpha} = B,\]
and hence
\[\alpha = \Big(\sum_{i}{\sqrt{w_i}}\Big)^2 / B^2.\]

The hit probability of file $i$ then equals
\[h_i = {U'_i}^{-1}(\alpha) = \frac{\sqrt{w_i}}{\sqrt{\alpha}} = \frac{\sqrt{w_i}}{\sum_{j}{\sqrt{w_j}}}B.\]

The utility function defined above is known to yield minimum potential delay fairness. It was shown in~\cite{kelly98} that the TCP congestion control protocol
implements such a utility function.

\subsubsection{$\boldsymbol{\beta \rightarrow\infty}$}\hspace*{\fill} \\
With $\beta \rightarrow\infty$, maximizing the sum of the utilities corresponds to (see~\cite{mo00} for proof)
\[\max_{h_i}{\min_{i}{h_i}}.\]

This utility function does not comply with the rules mentioned in Section~\ref{sec:model} for utility functions, as it is not strictly concave.
However, it is easy to see that the above utility function yields
\[h_i = B/N, \quad \forall{i}.\]

The utility function defined here maximizes the minimum hit probability, and corresponds to the \emph{max-min fairness}. Note that using identical utility functions for all files resulted in similar hit probabilities as this case. A brief summary of the utility functions discussed here is given in Table~\ref{tbl:u}. 
\begin{table*}[]
\centering
\caption{$\beta$-fair utility functions family}
\begin{tabular}{ | c | c | c | c |}
\hline
$\beta$ & $\max{\sum_{i}{U_i(h_i)}}$ & $h_i$ & implication \\
\hline
  0 & $\max{\sum{w_i h_i}}$ & $h_i = 1, i\le B, h_i = 0, i \ge B+1$ & maximizing overall throughput \\
  1 & $\max{\sum{w_i \log{h_i}}}$ & $h_i = w_i B / \sum_{j}{w_j}$ & proportional fairness \\
  2 & $\max{-\sum{w_i / h_i}}$ & $h_i = \sqrt{w_i} B / \sum_{j}{\sqrt{w_j}}$ & minimize potential delay \\
  $\infty$ & $\max{\min{h_i}}$ & $h_i = B/N$ & max-min fairness \\
\hline
\end{tabular}
\label{tbl:u}
\end{table*}


\section{Reverse Engineering}
\label{sec:reverse}
In this section, we study the widely used replacement-based caching policies, FIFO and LRU, and show that their hit/miss behaviors can be duplicated in our framework through an appropriate choice of utility functions.
 
It was shown in~\cite{Nicaise14} that, with a proper choice of timer values, a TTL cache can generate the same statistical properties, \ie~same hit/miss probabilities, as FIFO and LRU caching policies. 
In implementing these caches, non-reset and reset TTL caches are used for FIFO and LRU, respectively, with $t_i=T, i=1,\ldots,N$ where $T$ denotes the \emph{characteristic time}~\cite{Che01} of these caches. For FIFO and LRU caches with Poisson arrivals the hit probabilities can be expressed as
$h_i = 1 - 1/(1+\lambda_iT)$ and $h_i = 1 - e^{-\lambda_i T}$, and $T$ is computed such that $\sum_{i}{h_i} = B$.
For example for the LRU policy $T$ is the unique solution to the fixed-point equation
\[\sum_{i=1}^{N}{\left(1 - e^{-\lambda_i T}\right)} = B.\]


In our framework, we see from~\eqref{eq:hu} that the file hit probabilities depend on the Lagrange multiplier $\alpha$ corresponding to the cache size constraint in~\eqref{eq:opt}.
This suggests a connection between $T$ and $\alpha$. Further note that the hit probabilities are increasing functions of $T$. On the other hand, since utility functions are concave and increasing, $h_i = {U'_i}^{-1}(\alpha)$ is a decreasing
function of $\alpha$. Hence, we can denote $T$ as a decreasing function of $\alpha$, \ie~$T = f(\alpha)$. 

Different choices of function $f(\cdot)$ would result in different utility functions for FIFO and LRU policies. 
However, if we impose the functional dependence $U_i(h_i) = \lambda_i U_0(h_i)$, then the equation $h_i = {U'_i}^{-1}(\alpha)$ yields
\[h_i = {U'_0}^{-1}(\alpha/\lambda_i).\]
From the expressions for the hit probabilities of the FIFO and LRU policies, we obtain $T = 1/\alpha$. In the remainder of the section, we use this to derive utility functions for the FIFO and LRU policies.

\subsection{FIFO}
The hit probability of file $i$ with request rate $\lambda_i$ in a FIFO cache with characteristic time $T$ is
\[h_i = 1 - \frac{1}{1 + \lambda_i T}.\]
Substituting this into~\eqref{eq:hu} and letting $T = 1/\alpha$ yields
\[{U'_i}^{-1}(\alpha) = 1 - \frac{1}{1 + \lambda_i / \alpha}.\]
Computing the inverse of ${U'_i}^{-1}(\cdot)$ yields
\[U'_i(h_i) = \frac{\lambda_i}{h_i} - \lambda_i,\]
and integration of the two sides of the above equation yields the utility function for the FIFO cache 
\[U_i(h_i) = \lambda_i (\log{h_i} - h_i).\]

\subsection{LRU}
Taking $h_i = 1 - e^{-\lambda_i T}$ for the LRU policy and letting ${T = 1/\alpha}$ yields
\[{U'_i}^{-1}(\alpha) = 1 - e^{-\lambda_i/\alpha},\]
which yields
\[U'_i(h_i) = \frac{-\lambda_i}{\log{(1-h_i)}}.\]
Integration of the two sides of the above equation yields the utility function for the LRU caching policy
\[U_i(h_i) = \lambda_i \text{li}(1-h_i),\]
where $\text{li}(\cdot)$ is the logarithmic integral function
\[\text{li}(x) = \int_0^x{\frac{\diff t}{\ln{t}}}.\]

It is easy to verify, using the approach explained in Section~\ref{sec:opt}, that the utility functions computed
above indeed yield the correct expressions for the hit probabilities of the FIFO and LRU caches.
We believe these utility functions are unique if restricted to be multiplicative in\footnote{We note that utility functions, defined in this context, are subject to affine transformations, \ie~$aU+b$ yields the same hit probabilities as $U$ for any constant $a>0$ and $b$.} $\lambda_i$.

\section{Online Algorithms}
\label{sec:online}
In Section~\ref{sec:opt}, we formulated utility-driven caching as a convex optimization problem either with a fixed or an elastic cache size. However, it is not feasible to solve the optimization problem offline and then
implement the optimal strategy. Moreover, the system parameters can change over time. Therefore, we need algorithms
that can be used to implement the optimal strategy and adapt to changes in the system by collecting limited information.
In this section, we develop such algorithms.

\subsection{Dual Solution}
\label{sec:dual}
The utility-driven caching formulated in~\eqref{eq:opt} is a convex optimization problem, and hence the optimal solution corresponds to solving the dual problem.
The Lagrange dual of the above problem is obtained by incorporating the constraints into the maximization by means of Lagrange multipliers
\begin{align*}
\text{minimize} \quad &D(\alpha, \boldsymbol{\nu}, \boldsymbol{\eta}) = \max_{h_i}\Bigg\{ \sum_{i=1}^{N}{U_i(h_i)} \\
&\qquad\qquad\qquad -\alpha\left[ \sum_{i=1}^{N}{h_i} - B\right] \\
&\qquad\qquad\qquad -\sum_{i=1}^{N}{\nu_i (h_i - 1)} + \sum_{i=1}^{N}{\eta_i h_i} \Bigg\}\\
\text{such that} \quad &\alpha \ge 0, \quad \boldsymbol{\nu}, \boldsymbol{\eta} \ge \mathbf{0}.
\end{align*}
Using timer based caching techniques for controlling the hit probabilities with $0 < t_i < \infty$ ensures that $0 < h_i < 1$, and hence we have $\nu_i = 0$ and $\eta_i = 0$. 

Here, we consider an algorithm based on the dual solution to the utility maximization problem~\eqref{eq:opt}. Recall that we can write the Lagrange dual of the utility maximization problem as
\[D(\alpha) = \max_{h_i}{\left\{ \sum_{i=1}^{N}{U_i(h_i)}-\alpha\left[ \sum_{i=1}^{N}{h_i} - B\right] \right\}},\]
and the dual problem can be written as
\[\min_{\alpha \ge 0}{D(\alpha)}.\]

A natural decentralized approach to consider for minimizing $D(\alpha)$ is to gradually move the decision variables towards the optimal point using the gradient descent algorithm.
The gradient can be easily computed as
\[\frac{\partial D(\alpha)}{\partial\alpha} = -\Big(\sum_{i}{h_i} - B \Big),\]
and since we are doing a gradient \emph{descent}, $\alpha$ should be updated according to the \emph{negative} of the gradient as
\[\alpha \gets \max{\Big\{0, \alpha + \gamma \Big( \sum_{i}{h_i} - B \Big)\Big\}},\]
where $\gamma > 0$ controls the step size at each iteration. Note that the KTT conditions require that $\alpha \ge 0$.

Based on the discussion in Section~\ref{sec:opt}, to satisfy the optimality condition we must have
\[U'_i(h_i) = \alpha,\]
or equivalently
\[h_i = {U'_i}^{-1}(\alpha).\]
The hit probabilities are then controlled based on the timer parameters $t_i$ which can be set according to~\eqref{eq:non_reset_t} and~\eqref{eq:reset_t} for non-reset and reset TTL caches.

Considering the hit probabilities as indicators of files residing in the cache, the expression $\sum_{i}{h_i}$ can be interpreted as the number of items currently in the cache, denoted here as $B_{curr}$. We can thus summarize the control algorithm for a reset TTL algorithm as
\begin{align}
\label{eq:dual_sol}
t_i &= -\frac{1}{\lambda_i} \log{\Big(1 - {U'_i}^{-1}(\alpha) \Big)}, \notag\\
\alpha &\gets \max{\{0, \alpha + \gamma ( B_{curr} - B )\}}.
\end{align}
We obtain an algorithm for a non-reset TTL cache by using the correct expression for $t_i$ in~\eqref{eq:non_reset_t}.

Let $\alpha^*$ denote the optimal value for $\alpha$. We show in Appendix~\ref{appn:dual} that $D(\alpha) - D(\alpha^*)$ is a Lyapunov function and the above algorithm converges to the optimal solution.

\subsection{Primal Solution}
We now consider an algorithm based on the optimization problem in~\eqref{eq:opt_soft} known as the \emph{primal} formulation.

Let $W(\mathbf{h})$ denote the objective function in~\eqref{eq:opt_soft} defined as
\[W(\mathbf{h}) = \sum_{i=1}^{N}{U_i(h_i)} - C(\sum_{i=1}^{N}{h_i} - B).\]
A natural approach to obtain the maximum value for $W(\mathbf{h})$ is to use the gradient ascent algorithm.
The basic idea behind the gradient ascent algorithm is to move the variables $h_i$ in the direction of the gradient
\[\frac{\partial W(\mathbf{h})}{\partial h_i} = U'_i(h_i) - C'(\sum_{i=1}^{N}{h_i} - B).\]
Since the hit probabilities are controlled by the TTL timers, we move $h_i$ towards the optimal point by updating timers $t_i$.
Let $\dot{h_i}$ denote the derivative of the hit probability $h_i$ with respect to time. Similarly, define $\dot{t_i}$ as the derivative of the timer parameter $t_i$
with respect to time. We have
\[\dot{h_i} = \frac{\partial h_i}{\partial t_i}\dot{t_i}.\]
From~\eqref{eq:hit_non_reset} and~\eqref{eq:hit_reset}, it is easy to confirm that $\partial h_i/\partial t_i > 0$ for non-reset and reset TTL caches.
Therefore, moving $t_i$ in the direction of the gradient, also moves $h_i$s in that direction.

By gradient ascent, the timer parameters should be updated according to
\[t_i \gets \max{\left\{0, t_i + k_i\Big[ U'_i(h_i) - C'(B_{curr} - B)  \Big]\right\}},\]
where $k_i > 0$ is the step-size parameter, and $\sum_{i=1}^{N}{h_i}$ has been replaced with $B_{curr}$ based on the same argument as in the dual solution.

Let $\mathbf{h}^*$ denote the optimal solution to~\eqref{eq:opt_soft}. We show in Appendix~\ref{appn:primal} that $W(\mathbf{h}^*) - W(\mathbf{h})$ is a Lyapunov function, and the above algorithm converges to the optimal solution.

\subsection{Primal-Dual Solution}
Here, we consider a third algorithm that combines elements of the previous two algorithms.
Consider the control algorithm
\begin{align*}
t_i &\gets \max{\{0,  t_i + k_i [ U'_i(h_i) - \alpha]\}}, \\
\alpha &\gets \max{\{0, \alpha + \gamma (B_{current} - B)\}}.
\end{align*}
Using Lyapunov techniques we show in Appendix~\ref{appn:primal_dual} that the above algorithm converges to the optimal solution.

Now, rather than updating the timer parameters according to the above rule explicitly based on the utility function, we can have update rules based on a cache hit or miss.
Consider the following differential equation
\begin{equation}
\label{eq:t}
\dot{t_i} = \delta_m(t_i, \alpha)(1 - h_i)\lambda_i - \delta_h(t_i, \alpha)h_i\lambda_i,
\end{equation}
where $\delta_m(t_i, \alpha)$ and $-\delta_h(t_i, \alpha)$ denote the change in $t_i$ upon a cache miss or hit for file $i$, respectively.
More specifically, the timer for file $i$ is increased by $\delta_m(t_i, \alpha)$ upon a cache miss, and decreased by $\delta_h(t_i, \alpha)$ on a cache hit.

The equilibrium for~\eqref{eq:t} happens when $\dot{t_i} = 0$, which solving for $h_i$ yields
\[h_i = \frac{\delta_m(t_i, \alpha)}{\delta_m(t_i, \alpha) + \delta_h(t_i, \alpha)}.\]
Comparing the above expression with $h_i = {U'_i}^{-1}(\alpha)$ suggests that
$\delta_m(t_i, \alpha)$ and $\delta_h(t_i, \alpha)$ can be set to achieve desired hit probabilities and caching policies.

Moreover, the differential equation~\eqref{eq:t} can be reorganized as
\[\dot{t_i} = h_i \lambda_i \Big(\delta_m(t_i, \alpha)/h_i - [\delta_m(t_i, \alpha) + \delta_h(t_i, \alpha)]\Big),\]
and to move $t_i$ in the direction of the gradient $U'_i(h_i) - \alpha$ a natural choice for the update functions can be
\[\delta_m(t_i, \alpha) = h_i U'_i(h_i), \text{ and } \delta_m(t_i, \alpha) + \delta_h(t_i, \alpha) = \alpha.\]

To implement proportional fairness for example, these functions can be set as
\begin{equation}
\label{eq:prop_pd}
\delta_m(t_i, \alpha) = \lambda_i, \text{ and } \delta_h(t_i, \alpha) = \alpha - \lambda_i.
\end{equation}

For the case of max-min fairness, recall from the discussion in Section~\ref{sec:opt_identical} that a utility function that is content agnostic, \ie\ $U_i(h) = U(h)$, results in a max-min fair resource allocation. Without loss of generality we can have $U_i(h_i) = \log{h_i}$. Thus, max-min fairness can be implemented by having
\[\delta_m(t_i, \alpha) = 1, \text{ and } \delta_h(t_i, \alpha) = \alpha - 1.\]
Note that with these functions, max-min fairness can be implemented without requiring knowledge about request arrival rates~$\lambda_i$, while the previous approaches require such knowledge.


\subsection{Estimation of $\lambda_i$}
\label{sec:estimate}
Computing the timer parameter $t_i$ in the algorithms discussed in this section requires knowing the request arrival rates for most of the policies.
Estimation techniques can be used to approximate the request rates in case such knowledge is not available at the (cache) service provider.

Let $r_i$ denote the remaining TTL time for file $i$. Note that $r_i$ can be computed based on $t_i$ and a time-stamp for the last time file~$i$ was requested.
Let $X_i$ denote the random variable corresponding to the inter-arrival times for the requests for file~$i$, and $\bar{X_i}$ be its mean.
We can approximate the mean inter-arrival time as $\hat{\bar{X_i}} = t_i - r_i$. Note that $\hat{\bar{X_i}}$ defined in this way is a one-sample
unbiased estimator of $\bar{X_i}$. Therefore, $\hat{\bar{X_i}}$ is an unbiased estimator of $1/\lambda_i$. In the simulation section, we will use this estimator in computing the timer parameters for evaluating our algorithms.

\begin{figure*}[t]
\centering
 \begin{subfigure}[b]{0.25\linewidth}
  	\centering\includegraphics[scale=0.21]{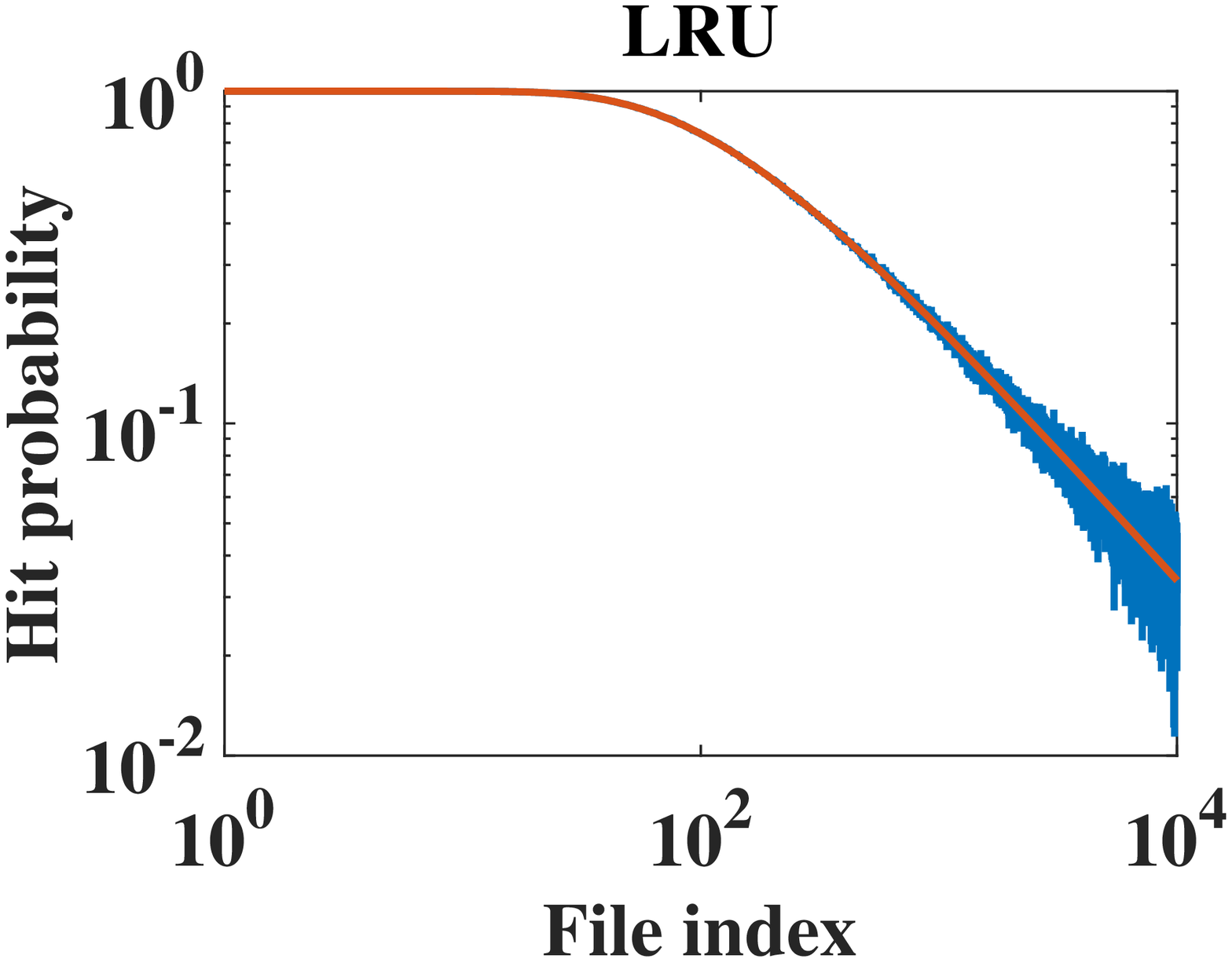}
 \end{subfigure}%
 \begin{subfigure}[b]{0.25\linewidth}
  	\centering\includegraphics[scale=0.21]{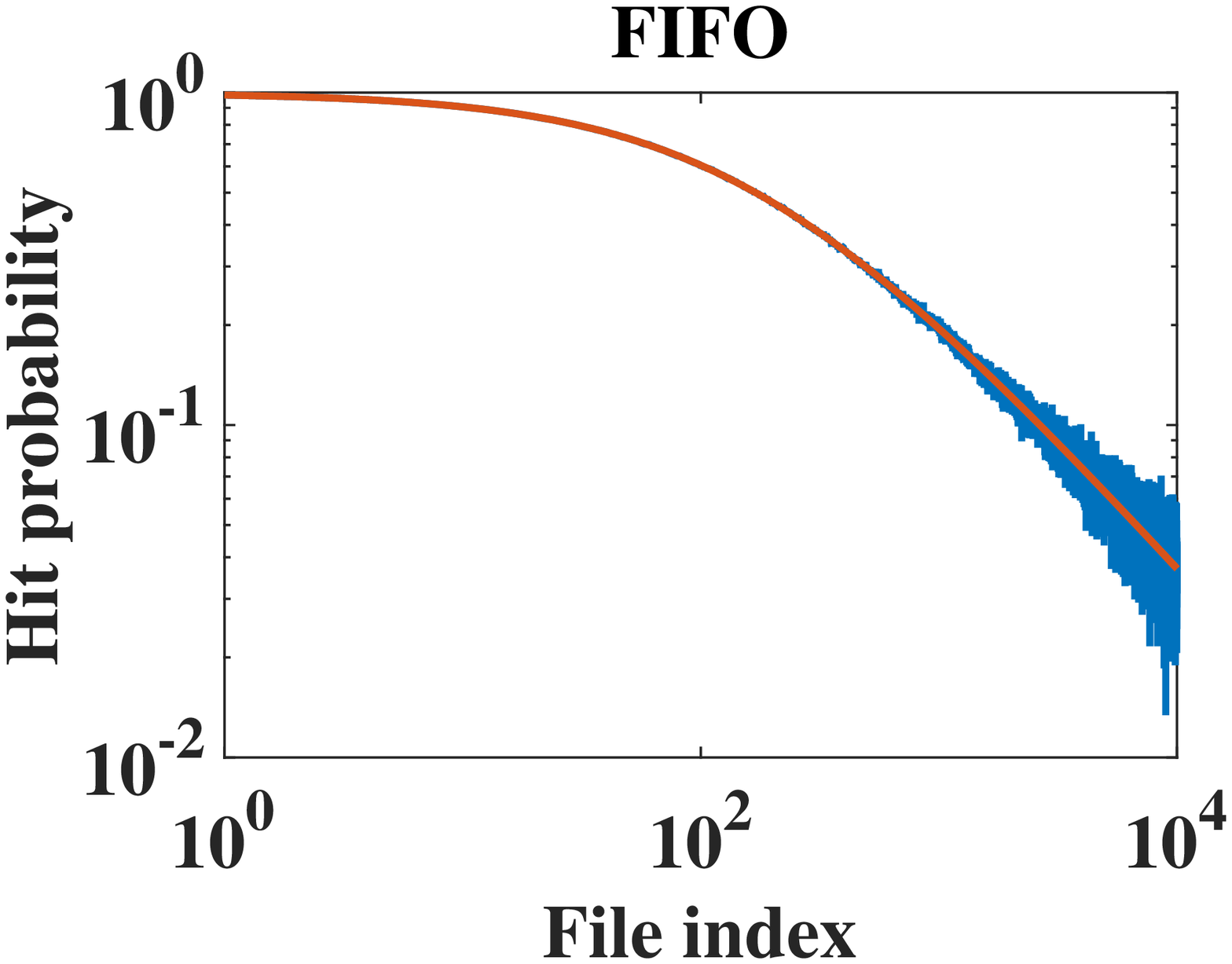}
 \end{subfigure}%
 \begin{subfigure}[b]{0.25\linewidth}
  	\centering\includegraphics[scale=0.21]{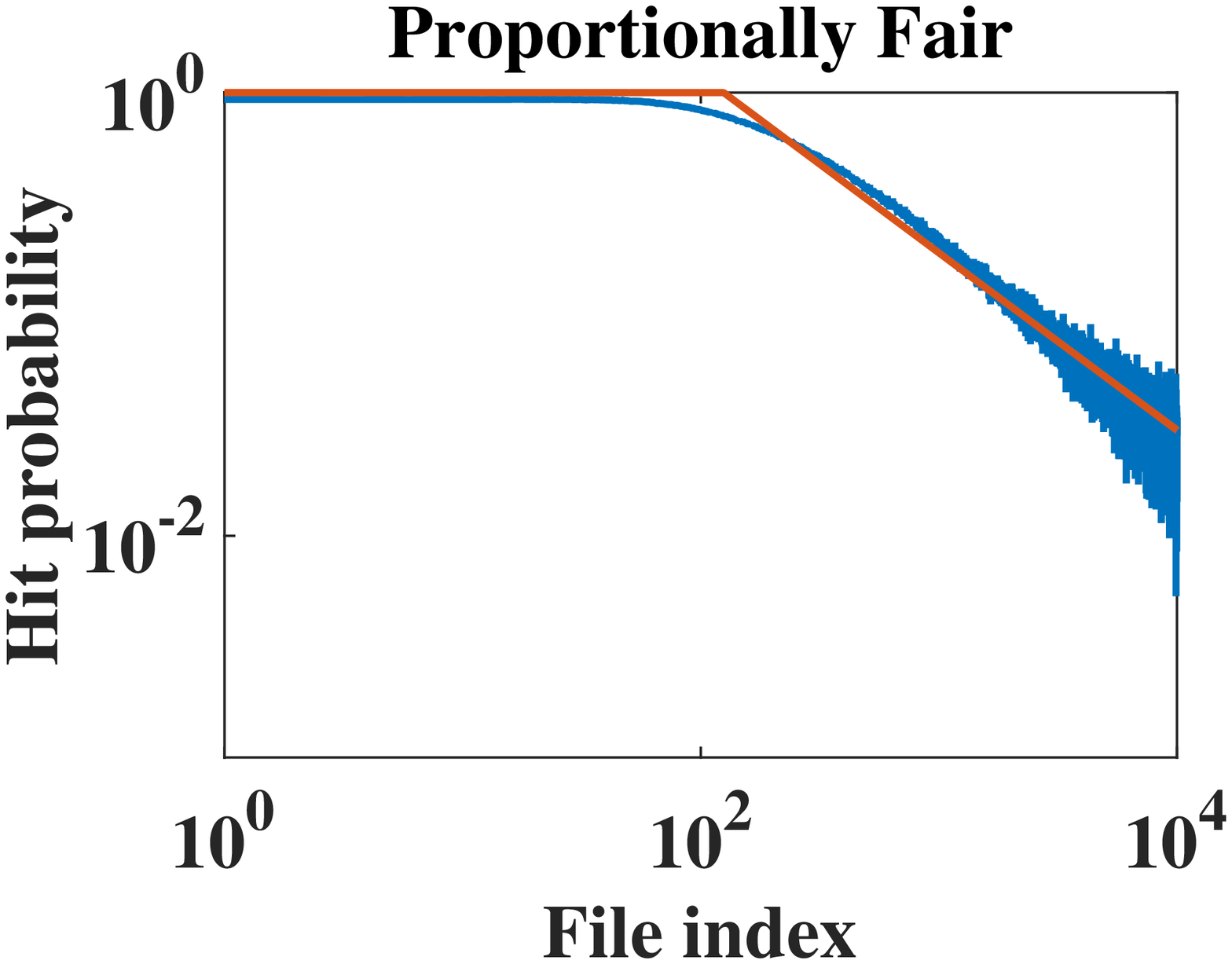}
 \end{subfigure}%
 \begin{subfigure}[b]{0.25\linewidth}
  	\centering\includegraphics[scale=0.21]{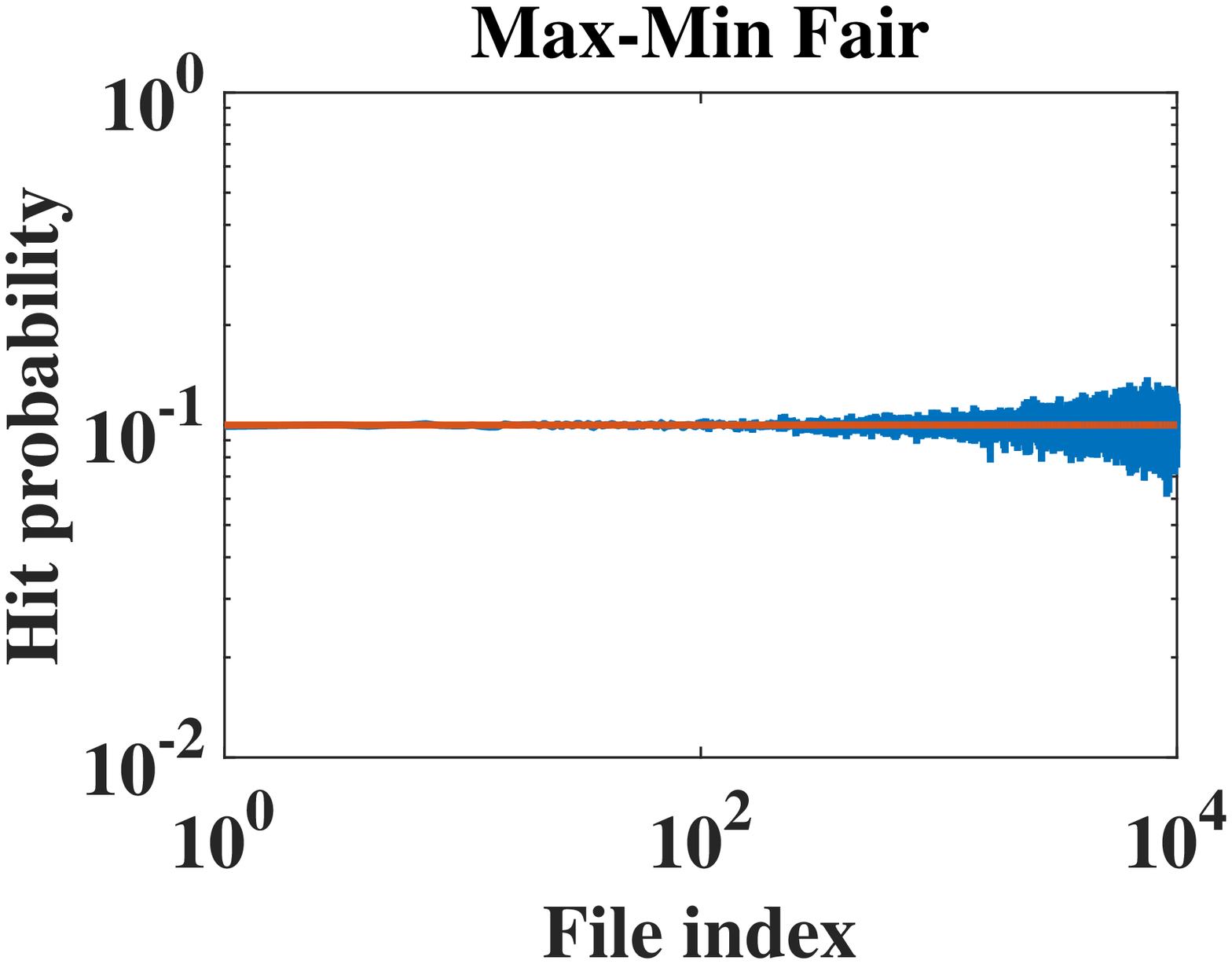}
 \end{subfigure}
\vspace{-0.25cm}
 \caption{Hit probabilities from implementing the online dual algorithm using utility functions for LRU, FIFO, proportionally fair and max-min fair policies.}
    \centering\label{fig:dual}
\end{figure*}

\section{Simulations}
\label{sec:simulation}
In this section, we evaluate the efficiency of the online algorithms developed in this paper.
Due to space restrictions, we limit our study to four caching policies: FIFO, LRU, proportionally fair, and max-min fair.

Per our discussion in Section~\ref{sec:reverse}, non-reset and reset TTL caches can be used with $t_i = T, i=1,\ldots,N$ to implement caches with the same statistical properties as FIFO and LRU caches.
However, previous approaches require precomputing the cache characteristic time $T$.
By using the online dual algorithm developed in Section~\ref{sec:dual} we are able to implement these policies with no a priori information of $T$.
We do so by implementing non-reset and reset TTL caches, with the timer parameters for all files
set as $t_i = 1/\alpha$, where $\alpha$ denotes the dual variable and is updated according to~\eqref{eq:dual_sol}.

For the proportionally fair policy, timer parameters are set to
\[t_i = \frac{-1}{\lambda_i}\log{(1 - \frac{\lambda_i}{\alpha})},\]
and for the max-min fair policy we set the timers as
\[t_i = \frac{-1}{\lambda_i}\log{(1 - \frac{1}{\alpha})}.\]
We implement the proportionally fair and max-min fair policies as reset TTL caches.


In the experiments to follow, we consider a cache with the expected number of files in the cache set to $B=1000$. Requests arrive for ${N = 10^4}$ files according to a Poisson process with aggregate rate one. File popularities follow a Zipf distribution with parameter ${s=0.8}$,~\ie~${\lambda_i = 1/i^s}$. In computing the timer parameters we use estimated values for the file request rates as explained in Section~\ref{sec:estimate}.

Figure~\ref{fig:dual} compares the hit probabilities achieved by our online dual algorithm with those computed numerically for the four policies explained above.
It is clear that the online algorithms yield the exact hit probabilities for the FIFO, LRU and max-min fair policies. For the proportionally fair policy however, the
simulated hit probabilities do not exactly match numerically computed values. This is due to an error in estimating $\lambda_i, i=1,\ldots, N$. Note that we use a simple estimator
here that is unbiased for $1/\lambda_i$ but biased for $\lambda_i$. It is clear from the above equations that computing timer parameters for the max-min fair policy
only require estimates of $1/\lambda_i$ and hence the results are good. Proportionally fair policy on the other hand requires estimating $\lambda_i$ as well,
hence using a biased estimate of $\lambda_i$ introduces some error.

\begin{figure}[h]
\centering
 \begin{subfigure}[b]{0.50\linewidth}
  	\centering\includegraphics[scale=0.20]{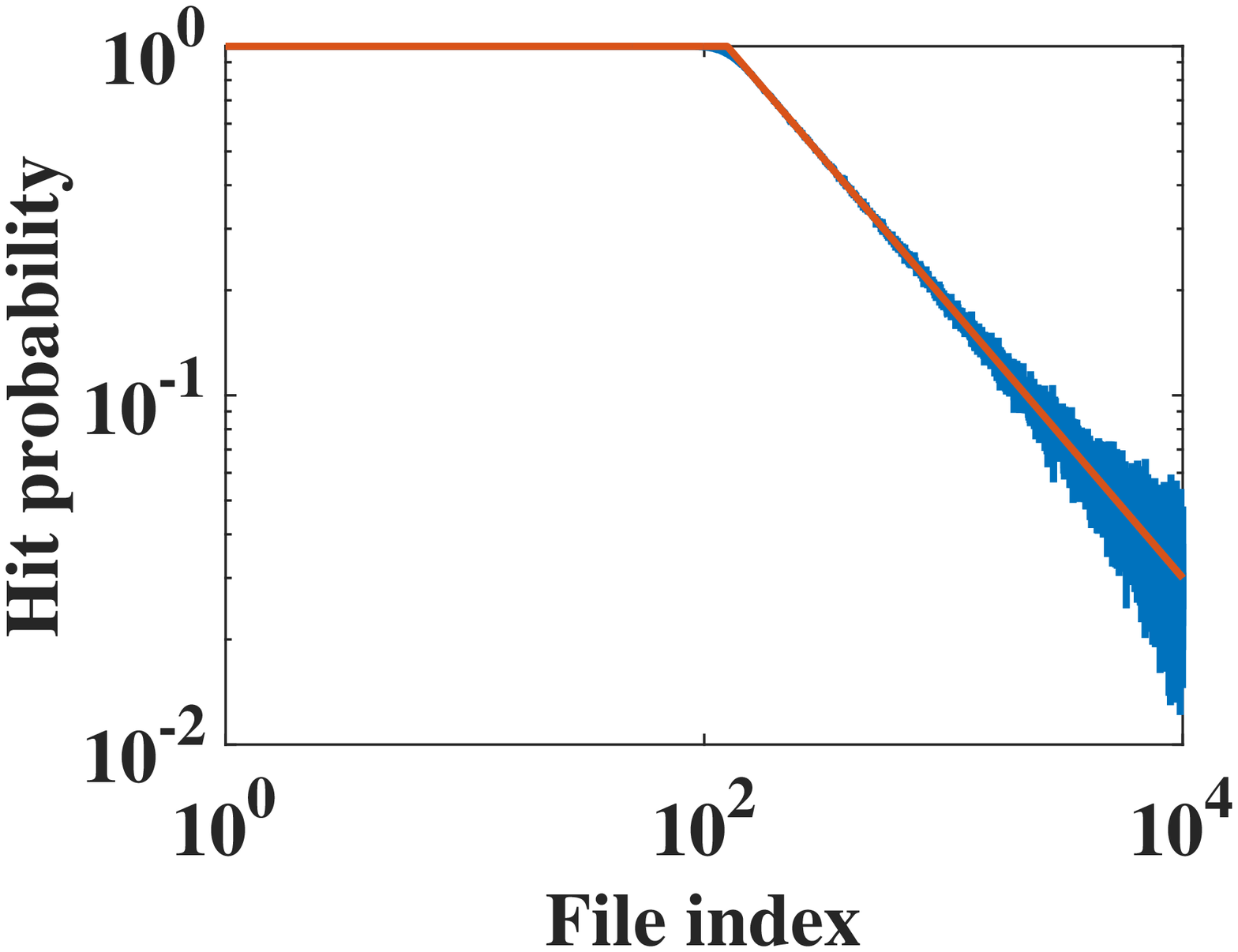}
  	\caption{\label{fig:prop_exact}}
 \end{subfigure}%
 \begin{subfigure}[b]{0.50\linewidth}
  	\centering\includegraphics[scale=0.20]{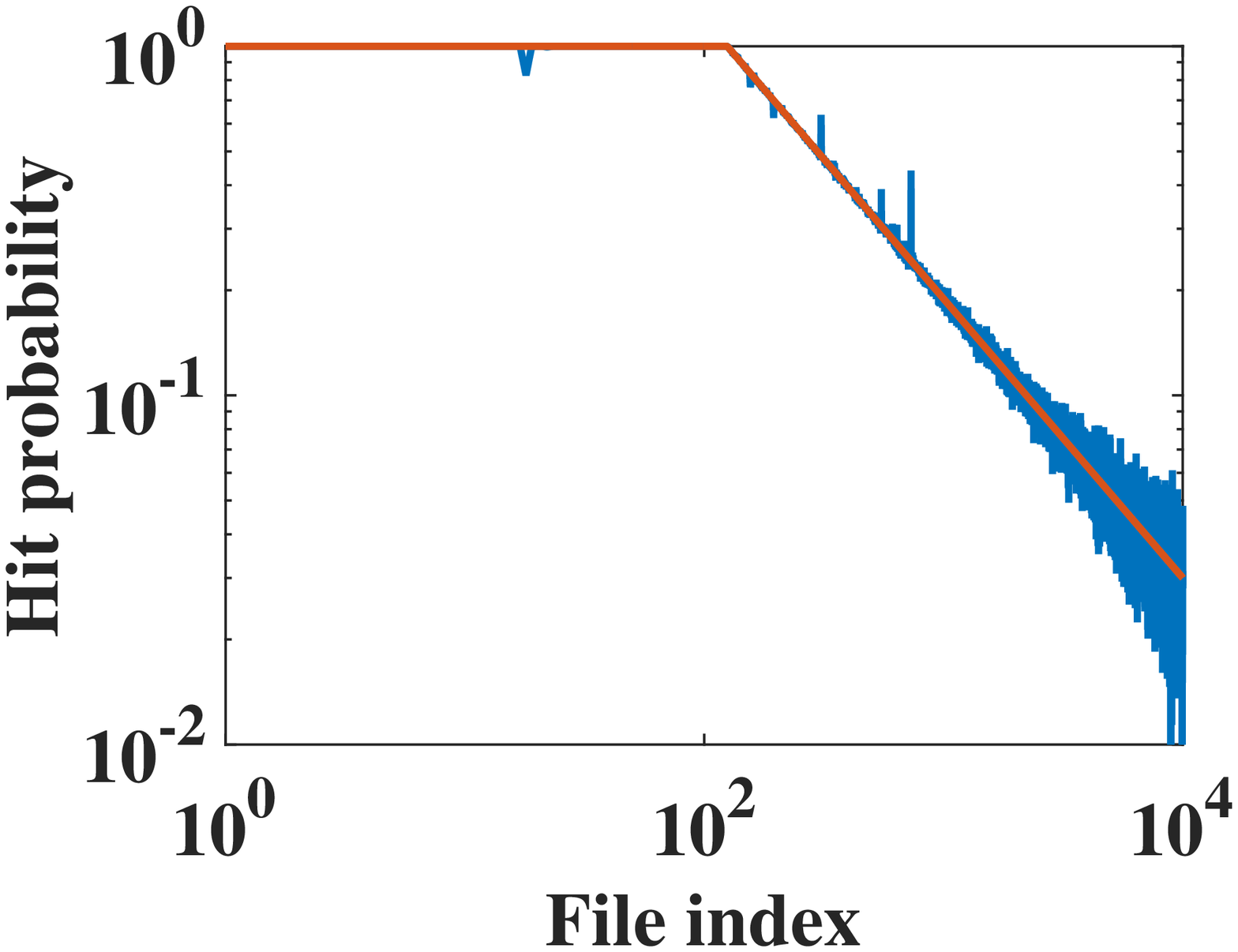}
  	\caption{\label{fig:prop_pd}}
 \end{subfigure}
\vspace{-0.25cm}
 \caption{Proportionally fair policy implemented using the (a) dual algorithm with exact knowledge of $\lambda_i$s, and (b) primal-dual algorithm with ${\delta_m(t_i, \alpha) = \lambda_i}$ and ${\delta_h(t_i, \alpha) = \alpha - \lambda_i}$, with approximate $\lambda_i$ values.}
  \label{fig:prop_fair}
\end{figure}

To confirm the above reasoning, we also simulate the proportionally fair policy assuming perfect knowledge of the request rates.
Figure~\ref{fig:prop_exact} shows that in this case simulation results exactly match the numerical values.

We can also use the primal-dual algorithm to implement the proportionally fair policy.
Here, we implement this policy using the update rules in~\eqref{eq:prop_pd}, and estimated values for the request rates.
Figure~\ref{fig:prop_pd} shows that, unlike the dual approach, the simulation results match the numerical values.
This example demonstrates how one algorithm may be more desirable than others in implementing a specific policy.

\begin{figure}[h]
\centering
 \begin{subfigure}[b]{0.5\linewidth}
  	\centering\includegraphics[scale=0.21]{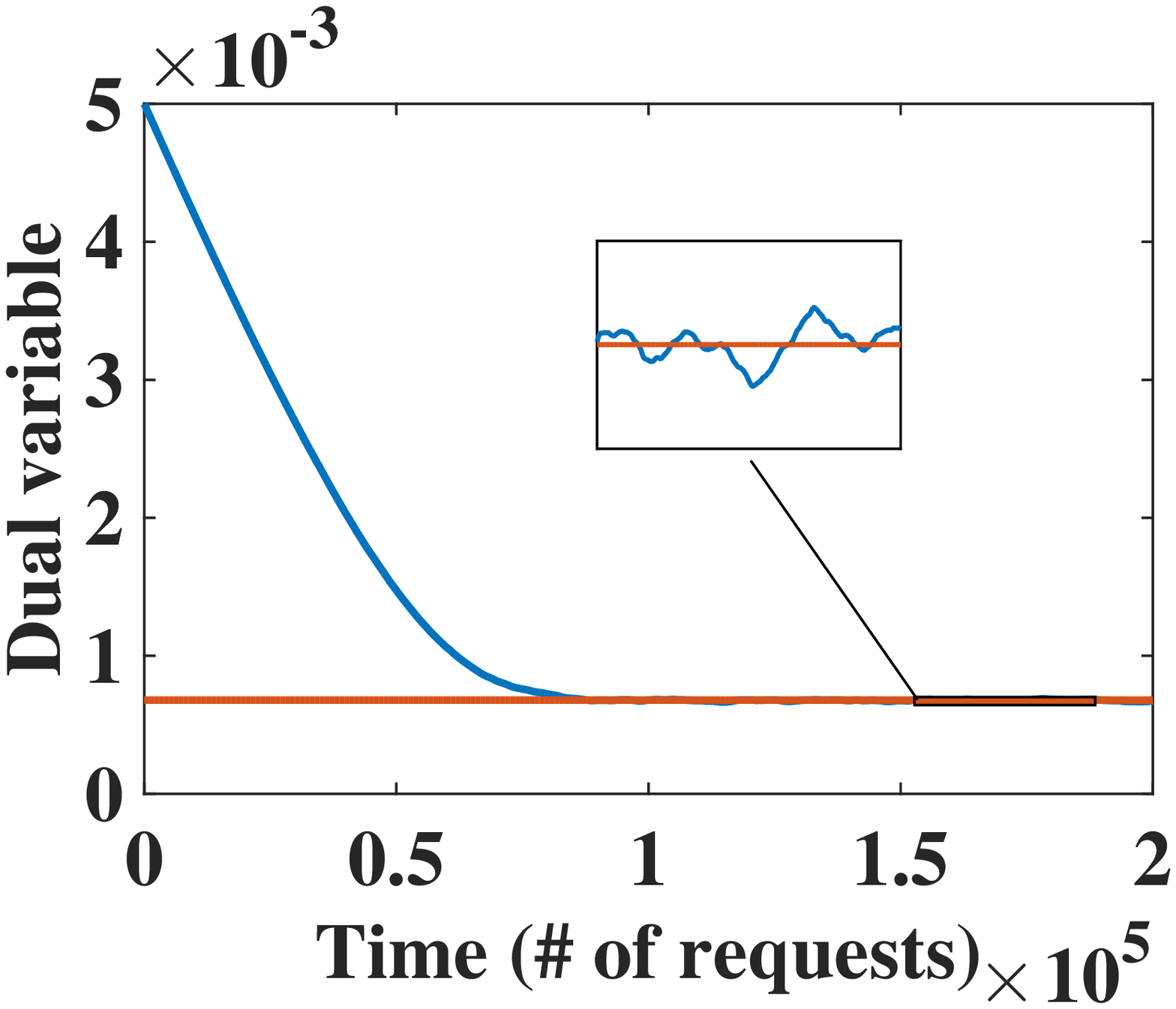}
  	\caption{\label{fig:lru_dual_var}}
 \end{subfigure}%
 \begin{subfigure}[b]{0.5\linewidth}
  	\centering\includegraphics[scale=0.21]{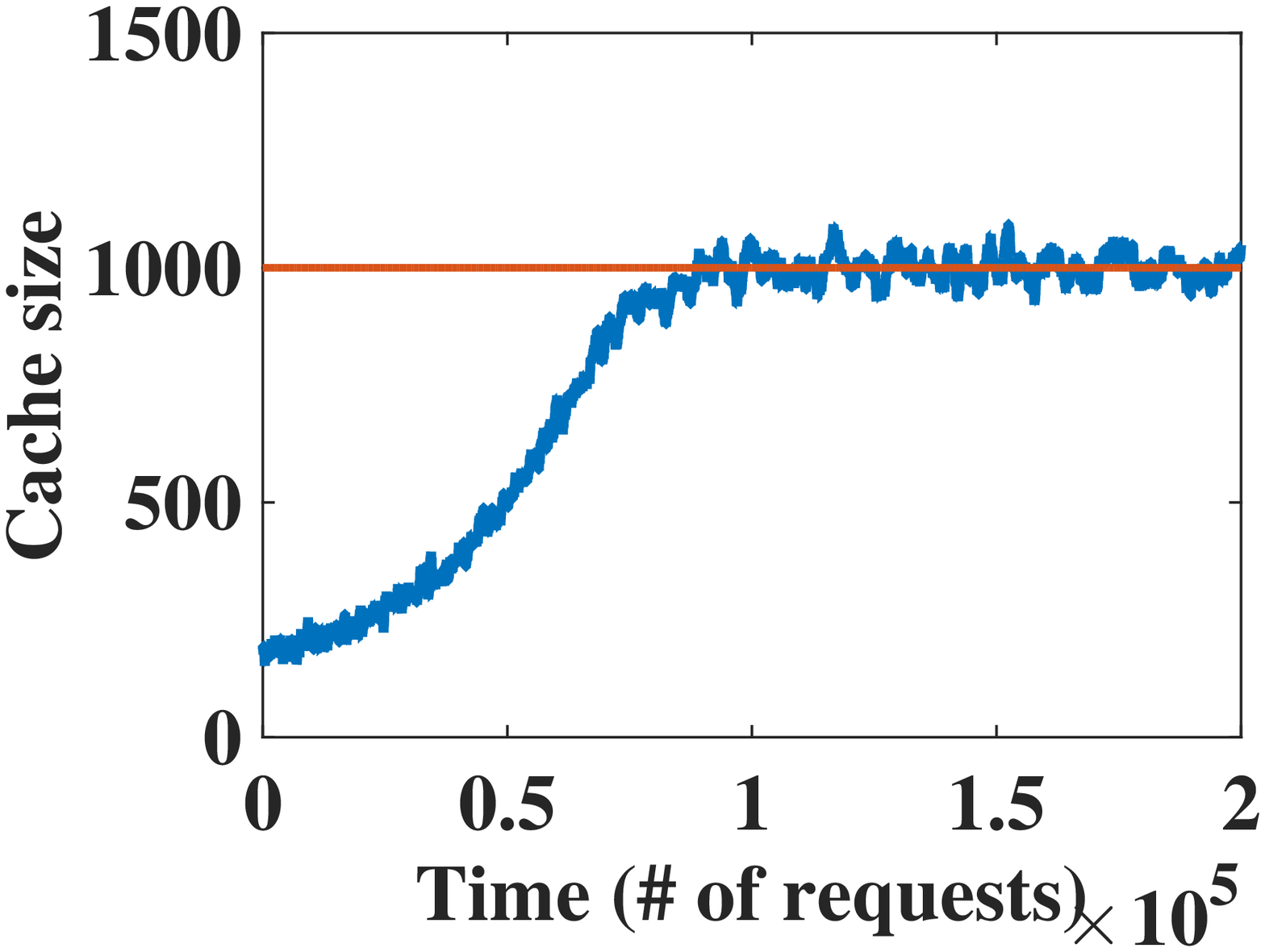}
  	\caption{\label{fig:lru_cache_size}}
 \end{subfigure}
\vspace{-0.25cm}
 \caption{Convergence and stability of dual algorithm for the utility function representing LRU policy.}
    \label{fig:lru_dual}
\end{figure}

\begin{figure}[h]
\centering
 \begin{subfigure}[b]{0.5\linewidth}
  	\includegraphics[scale=0.21]{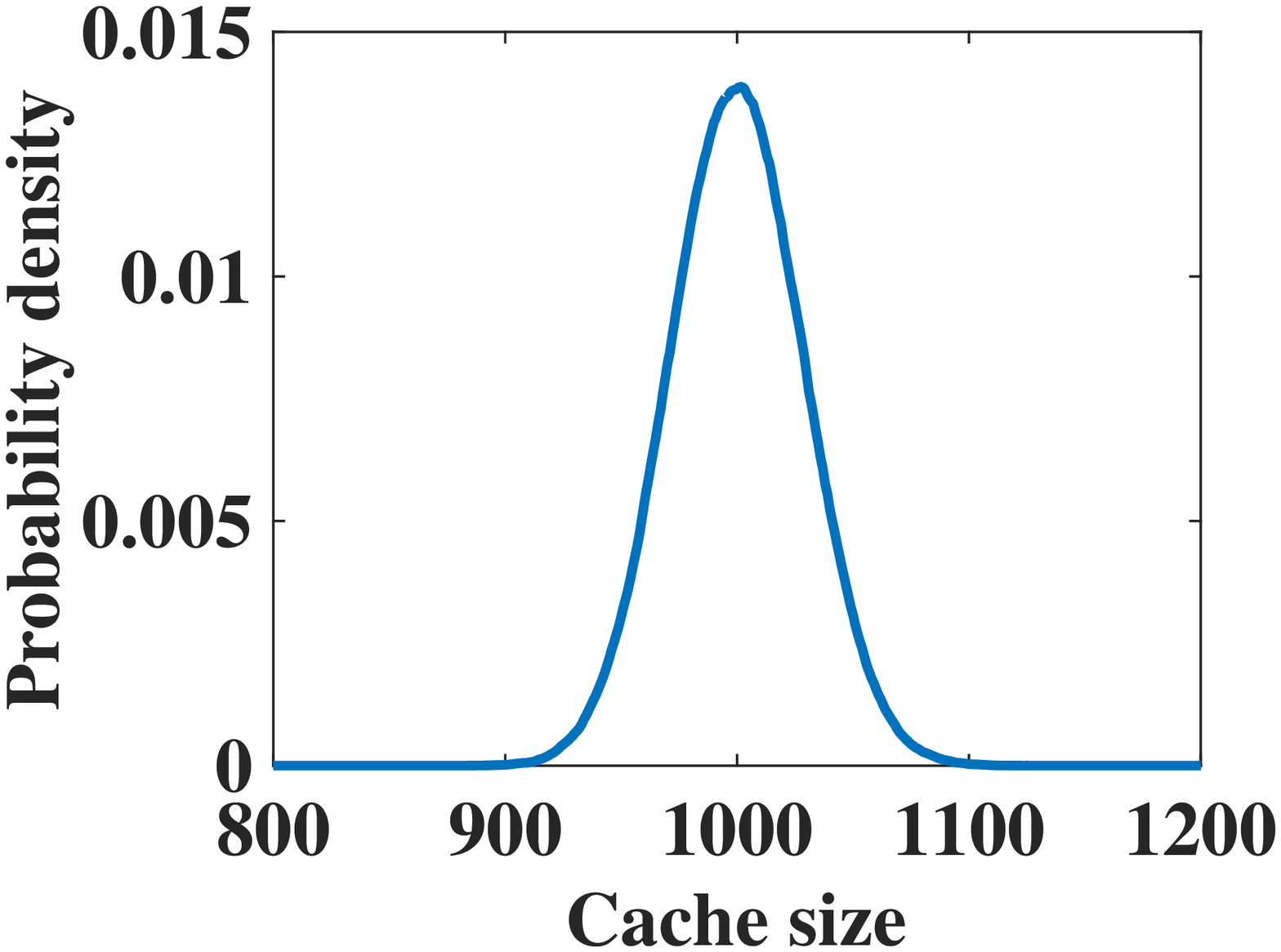}
 \end{subfigure}%
 \begin{subfigure}[b]{0.5\linewidth}
  	\includegraphics[scale=0.21]{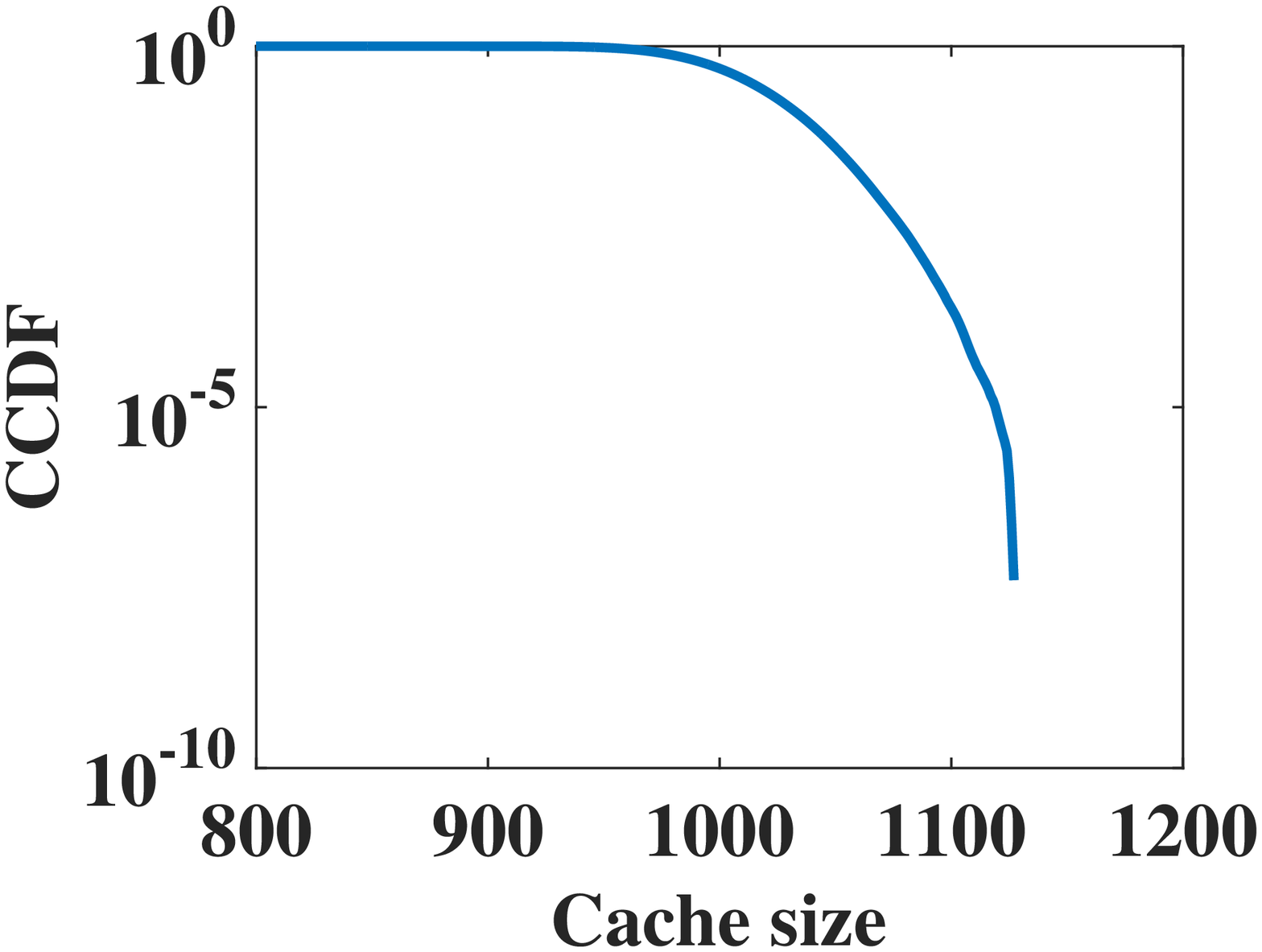}
 \end{subfigure}
\vspace{-0.25cm}
 \caption{Cache size distribution and CCDF from dual algorithm with the utility function representing LRU policy.}
    \label{fig:cs}
\end{figure}

The algorithms explained in Section~\ref{sec:online} are proven to be globally and asymptotically stable, and converge to the optimal solution.
Figure~\ref{fig:lru_dual_var} shows the convergence of the dual variable for the LRU policy.
The red line in this figure shows $1/T=6.8\times 10^{-4}$ where $T$ is the characteristic time of the LRU cache computed according to the discussion in Section~\ref{sec:reverse}.
Also, Figure~\ref{fig:lru_cache_size} shows how the number of contents in the cache is centered around the capacity $B$.
The probability density and complementary cumulative distribution function (CCDF) for the number of files in cache are shown in Figure~\ref{fig:cs}.
The probability of violating the capacity $B$ by more than $10\%$ is less than $2.5\times 10^{-4}$. For larger systems, \ie\ for large $B$ and $N$, the probability of violating the 
target cache capacity becomes infinitesimally small; see the discussion in Section~\ref{sec:violation}. This is what we also observe in our simulations.
Similar behavior in the convergence of the dual variable and cache size is observed in implementing the other policies as well.

\section{Discussion}
\label{sec:discussion}
In this section, we explore the implications of utility-driven caching on monetizing the caching service and discuss some future research directions.

\subsection{Decomposition}
The formulation of the problem in Section~\ref{sec:opt} assumes that the utility functions $U_i(\cdot)$ are known to the system. In reality content providers might decide not to share their utility functions with the service provider. To handle this case, we decompose the optimization problem~\eqref{eq:opt} into two simpler problems.

Suppose that the cache storage is offered as a service and the service provider charges content providers at a constant rate $r$ for storage space. Hence, a content provider needs to pay an amount of $w_i = rh_i$ to receive hit probability $h_i$ for file $i$. The utility maximization problem for the content provider of file $i$ can then be written as
\begin{align}
\label{eq:opt_user}
\text{maximize} \quad &U_i(w_i/r) - w_i \\
\text{such that} \quad &w_i \ge 0 \notag
\end{align}

Now, assuming that the service provider knows the vector $\mathbf{w}$, for a proportionally fair resource allocation, the hit probabilities should be
set according to
\begin{align}
\label{eq:opt_network}
\text{maximize} \quad &\sum_{i=1}^{N}{w_i\log{(h_i)}} \\
\text{such that} \quad &\sum_{i=1}^{N}{h_i} = B \notag
\end{align}

It was shown in~\cite{kelly97} that there always exist vectors $\mathbf{w}$ and $\mathbf{h}$, such that $\mathbf{w}$ solves~\eqref{eq:opt_user} and $\mathbf{h}$ solves~\eqref{eq:opt_network}; further, the vector $\mathbf{h}$ is the unique solution to~\eqref{eq:opt}.

\subsection{Cost and Utility Functions}
In Section~\ref{sec:soft}, we defined a penalty function denoting the cost of using additional storage space. One might also define cost functions based on the consumed network bandwidth. This is especially interesting in modeling in-network caches with network links that are likely to be congested.

Optimization problem~\eqref{eq:opt} uses utility functions defined as functions of the hit probabilities. It is reasonable to define utility as a function of the hit \emph{rate}. Whether this makes any changes to the problem, \eg\ in the notion of fairness, is a question that requires further investigation. One argument in support of utilities as functions of hit rates is that a service provider might prefer pricing based on request rate rather than cache occupancy. Moreover, in designing hierarchical caches a service provider's objective could be to minimize the internal bandwidth cost. This can be achieved by defining the utility functions as $U_i = -C_i(m_i)$ where $C_i(m_i)$ denotes the cost associated with miss rate $m_i$ for file $i$.

\subsection{Online Algorithms}
In Section~\ref{sec:online}, we developed three online algorithms that can be used to implement utility-driven caching. Although these algorithms are proven to be stable and converge to the optimal solution, they have distinct features that can make one algorithm more effective in implementing a policy. For example, implementing the max-min fair policy based on the dual solution requires knowing/estimating the file request rates, while it can be implemented using the modified primal-dual solution without such knowledge. Moreover, the convergence rate of these algorithms may differ for different policies. The choice of non-reset or reset TTL caches also has implications on the design and performance of these algorithms.
These are subjects that require further study.

\section{Conclusion}
\label{sec:conclusion}
In this paper, we proposed the concept of utility-driven caching, and formulated it as an optimization problem with rigid and elastic cache storage size constraints. Utility-driven caching provides a general framework for defining caching policies with considerations of fairness among various groups of files, and implications on market economy for (cache) service providers and content publishers. This framework has the capability to model existing caching policies such as FIFO and LRU, as utility-driven caching policies.

We developed three decentralized algorithms that implement utility-driven caching policies in an online fashion and that can adapt to changes in file request rates over time. We prove that these algorithms are globally stable and converge to the optimal solution. Through simulations we illustrated the efficiency of these algorithms and the flexibility of our approach.

\bibliographystyle{IEEEtran}
\bibliography{references}

\begin{appendices}

\section{Stability of Dual Solution}
\label{appn:dual}
We first note that $D(\alpha)$ is the dual of a convex function and has a unique minimizer $\alpha^*$.
The function ${V(\alpha) = D(\alpha) - D(\alpha^*)}$, hence, is a non-negative function and equals zero only at $\alpha = \alpha^*$.
Differentiating $V(\alpha)$ with respect to time we get
\[\dot{V}(\alpha) = \frac{\partial V}{\partial \alpha}\dot{\alpha} = -\gamma \Big(\sum_{i}{h_i} - B \Big)^2 < 0.\]

Therefore, $V(\cdot)$ is a Lyapunov function, and the system state will converge to optimum starting from any initial condition.

\section{Stability of Primal Solution}
\label{appn:primal}
We first note that since $W(\h)$ is a strictly concave function, it has a unique maximizer $\h^*$.
Moreover ${V(\h) = W(\h^*) - W(\h)}$ is a non-negative function and equals zero only at $\h = \h^*$.
Differentiating $V(\cdot)$ with respect to time we obtain
\begin{align*}
\dot{V}(\h) &= \sum_{i}{\frac{\partial V}{\partial h_i}\dot{h_i}} \\
&= -\sum_{i}{\left( U'_i(h_i) - C'(\sum_{i}{h_i} - B) \right) \dot{h_i}}.
\end{align*}

For $\dot{h_i}$ we have
\[\dot{h_i} = \frac{\partial h_i}{\partial t_i}\dot{t_i}.\]
For non-reset and reset TTL caches we have
\[\frac{\partial h_i}{\partial t_i} = \frac{\lambda_i}{(1+\lambda_i t_i)^2} \qquad\text{ and }\qquad \frac{\partial h_i}{\partial t_i} = \lambda_i e^{-\lambda_i t_i},\]
respectively, and hence $\partial h_i / \partial t_i > 0$.

From the controller for $t_i$ we have
\[t_i = k_i \left( U'_i(h_i) - C'(\sum_{i}{h_i} - B) \right).\]

Hence, we get
\[\dot{V}(\h) = -\sum_{i}{k_i \frac{\partial h_i}{\partial t_i} \left( U'_i(h_i) - C'(\sum_{i}{h_i} - B) \right)^2} < 0.\]

Therefore, $V(\cdot)$ is a Lyapunov function\footnote{A description of Lyapunov functions and their applications can be found in~\cite{srikant13}.}, and the system state will converge to $\h^*$ starting from any initial condition.

\section{Stability of Primal-Dual Solution}
\label{appn:primal_dual}
As discussed in Section~\ref{sec:opt}, the Lagrangian function for the optimization problem~\eqref{eq:opt} is expressed as
\[\LL(\h, \alpha) = \sum_{i}{U_i(h_i)} - \alpha(\sum_{i}{h_i} - B).\]
Note that $\LL(\h, \alpha)$ is concave in $\h$ and convex in $\alpha$, and hence first order condition for optimality of $\h^*$ and $\alpha^*$ implies
\begin{align*}
\LL(\h^*, \alpha) &\le \LL(\h, \alpha) + \sum_{i}{\frac{\partial \LL}{\partial h_i}(h_i^* - h_i)}, \\
\LL(\h, \alpha^*) &\ge \LL(\h, \alpha) + \frac{\partial \LL}{\partial \alpha}(\alpha^* - \alpha) .
\end{align*}

Assume that the hit probability of a file can be expressed by $f(\cdot)$ as a function of the corresponding timer value $t_i$, \ie\ ${h_i = f(t_i)}$. The temporal derivative of the hit probability can therefore be expressed as
\[\dot{h_i} = f'(t_i) \dot{t_i},\]
or equivalently
\[\dot{h_i} = f'(f^{-1}(h_i)) \dot{t_i},\]
where $f^{-1}(\cdot)$ denotes the inverse of function $f(\cdot)$. For notation brevity we define ${g(h_i) = f'(f^{-1}(h_i))}$. Note that as discussed in Appendix~\ref{appn:primal}, $f(\cdot)$ is an increasing function, and hence ${g(h_i)\ge 0}$.

In the remaining, we show that $V(\h, \alpha)$ defined below is a Lyapunov function for the primal-dual algorithm:
\[V(\h, \alpha) = \sum_{i}{\int_{h_i^*}^{h_i}{\frac{x - h_i^*}{k_i g(x)}\diff x}} + \frac{1}{2\gamma}(\alpha - \alpha^*)^2.\]
Differentiating the above function with respect to time we obtain
\[\dot{V}(\h, \alpha) = \sum_{i}{\frac{h_i - h_i^*}{k_i g(h_i)}\dot{h_i}} + \frac{\alpha - \alpha^*}{\gamma}\dot{\alpha}.\]
Based on the controllers defined for $t_i$ and $\alpha$ we have
\[\dot{h_i} = g(h_i) \dot{t_i} = k_i g(h_i) \frac{\partial \LL}{\partial h_i},\]
and
\[\dot{\alpha} = -\gamma\frac{\partial \LL}{\partial \alpha}.\]
Replacing for $\dot{h_i}$ and $\dot{\alpha}$ in $\dot{V}(\h, \alpha)$, we obtain
\begin{align*}
\dot{V}(\h, \alpha) &= \sum_{i}{(h_i - h_i^*)\frac{\partial \LL}{\partial h_i}} - (\alpha - \alpha^*)\frac{\partial \LL}{\partial \alpha} \\
&\le \LL(\h, \alpha) - \LL(\h^*, \alpha) + \LL(\h, \alpha^*) - \LL(\h, \alpha) \\
&= \Big(\LL(\h^*, \alpha^*) - \LL(\h^*, \alpha)\Big) + \Big(\LL(\h, \alpha^*) - \LL(\h^*, \alpha^*)\Big) \\
&\le 0,
\end{align*}
where the last inequality follows from 
\[\LL(\h, \alpha^*)  \le \LL(\h^*, \alpha^*)  \le \LL(\h^*, \alpha),\]
for any $\h$ and $\alpha$.

Moreover, $V(\h, \alpha)$ is non-negative and equals zero only at $(\h^*, \alpha^*)$.
Therefore, $V(\h, \alpha)$ is a Lyapunov function, and the system state will converge to optimum starting from any initial condition.

\end{appendices}

\end{document}